\documentclass[showpacs,prl,twocolumn,superscriptaddress]{revtex4}
\usepackage{epsfig}
\usepackage{bm}

\begin{document}
\title{
Spin, Orbital and Charge Order at the Interface between Correlated Oxides
}
\author{G.~Jackeli}
\altaffiliation[]{Also at E.~Andronikashvili Institute of Physics,
0177 Tbilisi, Georgia} \affiliation{Max-Planck-Institut f\"ur
Festk\"orperforschung, Heisenbergstrasse 1, D-70569 Stuttgart,
Germany}
\author{G.~Khaliullin}
\affiliation{Max-Planck-Institut f\"ur Festk\"orperforschung,
Heisenbergstrasse 1, D-70569 Stuttgart, Germany}

\begin{abstract}
The collective behavior of correlated electrons in the VO$_2-$interface layer
of LaVO$_3$/SrTiO$_3$ heterostructure is studied within a quarter-filled
$t_{2g}$-orbital Hubbard model on a square lattice.
We argue that the ground state is ferromagnetic driven
by the double exchange mechanism, and is orbitally and
charge ordered due to a confined geometry and electron correlations.
The orbital and charge density waves open gaps on the entire Fermi surfaces
of all orbitals. The theory explains the observed insulating behavior of
the $p$-type interface between LaVO$_3$ and SrTiO$_3$.
\end{abstract}
\maketitle

The recent progress in manufacturing and experimental studies of 
heterostructures and superlattices based on the transition metal 
oxides has lead to the discoveries of a number of novel physical 
phenomena and new electronic states emerging at the interfaces 
\cite{Ued98,Oht02,Yam04,Oht04,Tsu07,Rey07,Bri07,Cha06,Cha07}. 
A formation of a high-mobility electron gas \cite{Oht04}, quantum Hall
effect \cite{Tsu07} and, remarkably
enough, even a superconducting state \cite{Rey07} have been observed at the
interfaces between insulating oxides. At the interface between nominally
non-magnetic oxides, magnetic effects have also been detected \cite{Bri07}.
The physical properties of interfaces may largely differ from and
can even be orthogonal to those of bulk materials, due to
an "electronic reconstruction" phenomenon \cite{Oka04}.

In correlated oxide heterostructures, electronic reconstruction involves
not only charge, but also spin \cite{Cha06} and, in particular, the orbital
degrees of freedom \cite{Cha07,Dag07,Cha08} since electronic orbitals are
highly sensitive to the local environment.
Such a multifaceted response of correlated electrons gives rise to the
rich interface physics that may form a basis for future device applications.

The recent work \cite{Hot07} added a new puzzle into this field.
Two types of interfaces between a Mott insulator LaVO$_3$ and
a band insulator SrTiO$_3$ have been investigated:
(1) the VO$_2$/LaO/$TiO_2$/SrO interface with $n$-type polar
discontinuity, and (2) the LaO/$VO_2$/SrO/TiO$_2$ with $p$-type one
(formed by inserting a "metallic" SrVO$_3$ unit into the $n$-type interface).
In bulk compounds, the VO$_2$ (LaO) layers have a $-e$ ($+e$) charge
per unit cell, while TiO$_2$ and SrO layers are neutral.
In such systems a polar discontinuity
triggers the doping of an interface layer to resolve the polar catastrophe
\cite{Nak06}: In $n$-type interface the TiO$_2$ layer receives a $-e/2$ charge
while in $p$-type interface $-e/2$ charge is taken away from the VO$_2$
layer. This leads to a formal valence state $d^{0.5}$ of Ti and $d^{1.5}$ of V
at the $n$- and $p$-type interface layers, respectively. The resistivity
measurements have shown that the $n$-type interface is metallic and $p$-type
interface is insulating. The metallic character of TiO$_2$ interface
layer is not surprising and confirms existing theoretical
results \cite{Oka04}. However, an insulating behavior of hole doped VO$_2$
interface layer is at odds with expectations and is striking,
given that SrVO$_3$ is a good metal and already $18\%$ Sr-doping is
sufficient to convert bulk LaVO$_3$ into a metal, too \cite{Miy00}.
In this Letter, we present a theory resolving this puzzle.
In short, the contrasting behavior of TiO$_2$ and VO$_2$ interfaces
originates from their different spin and orbital structure.
Indeed, while TiO$_2$ layer with Ti$^{3+}$/Ti$^{4+}$ states
represents a diluted quantum S=1/2 system (like high-T$_c$ cuprates),
VO$_2$ interface layer is made of V$^{3+}$ S=1 and V$^{4+}$ S=1/2 states --
a canonical background for the double-exchange (DE) physics. Once spins of
the VO$_2$ layer are polarized by the DE mechanism,
system is effectively {\it half-filled} and hence collective orbital
and charge instabilities are triggered at the interface. We argue that
these cooperative orderings of correlated electrons are
responsible for the insulating character of the $p$-type interface.

\begin{figure}
\epsfysize=60mm \centerline{\epsffile{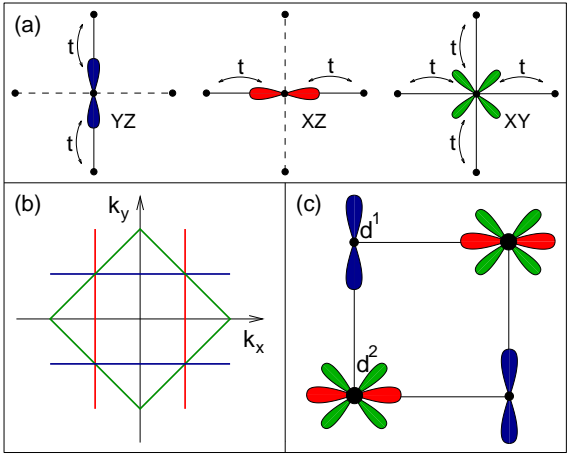}} \caption{(Color
online) (a) The hopping amplitudes of $t_{2g}$-electrons on a square
lattice. (b) The Fermi surfaces for $d_{yz}$ (horizontal lines),
$d_{xz}$ (vertical lines) and $d_{xy}$ (square) electrons in the
ferromagnetic state. (c) A sketch of the orbital and charge ordered
state for $U\!>\!V\!\gg\!t$. In a symmetry broken state the system
is insulating at any values of $U$ and $V$.} \label{fig1}
\end{figure}

{\it The Model.}-- We describe the physics of hole doped VO$_2$ layer
within a multi-orbital Hubbard model for $d$-electrons \cite{Cas78}
on a square lattice:
\begin{eqnarray}
\label{h}
H\!\!&=&\!\!- \sum_{i,j}\sum_{\alpha,\sigma}\!t_{ij,\alpha}
d_{i\alpha\sigma}^{\dagger}{d}_{j\alpha\sigma} \;+\;
U\sum_{i,\alpha}n_{i\alpha\uparrow}\!n_{i\alpha\downarrow}\\
&+&\!\!\!
\sum_{i,\alpha<\beta}\!\left[U'-2J_{\rm H}({\vec s}_{i\alpha}{\vec s}_{i\beta}
+\frac{1}{4})\right]\!n_{i\alpha}n_{i\beta}+V\sum_{\langle ij\rangle}n_in_j.
\nonumber
\end{eqnarray}
The three-fold degenerate $t_{2g}$ states $d_{yz}$, $d_{xz}$, and
$d_{xy}$ are labeled by orbital index $\alpha=$1, 2, and 3,
respectively. The ${\vec s}_{i\alpha}$ and
$n_{i\alpha}=n_{i\alpha\uparrow}+n_{i\alpha\downarrow}$ correspond
to the spin and density of electrons in $\alpha$ orbital. The first
term in $H$ describes an electron hopping between the nearest
neighbor (NN) sites \cite{Note0} and is diagonal in orbital space.
The peculiarity of a $t_{2g}$ system on a planar geometry is that
the orbitals $d_{yz}$ and $d_{xz}$ become one-dimensional (1D). 
They have a finite hopping amplitude only along the one
particular direction, $t_{ij,1(2)}=t$ for $ij\parallel x(y)$ and
zero otherwise, see Fig.~\ref{fig1}(a). While the $d_{xy}$ orbital
still forms a two-dimensional (2D) band: $t_{ij,3}=t$. In momentum
space the hopping term reads as 
$\sum_{{\bf k}\alpha\sigma}\epsilon_{{\bf k} \alpha}d_{{\bf
k}\alpha\sigma}^{\dagger}{d}_{{\bf k}\alpha\sigma}$, where
$\epsilon_{{\bf k}1(2)}=-2t\cos k_{x(y)}$ and $\epsilon_{{\bf k}3}=
-2t(\cos k_{x}+\cos k_{y})$. The interaction part of $H$ consists of an
on-site intra- and inter-orbital Coulomb repulsions, $U$ and $U'$,
respectively. The latter term is further split by the Hund's
coupling $J_{\rm H}$ into an interorbital spin triplet ($U'-J_{\rm
H}$) and singlets ($U'+J_{\rm H}$), such that $S=1$ state with
electrons residing on different orbitals is favored. It is this
$2J_{\rm H}$ splitting between the high/low spin states that
promotes a global ferromagnetic (FM) state by virtue of the DE
mechanism in the present $V^{3+/4+}$ mixed-valent system. We have
also included NN repulsion $V$ which is relevant for charge ordering
\cite{Note1}. We consider a quarter-filled $t_{2g}$ bands
($d^{1.5}$ configuration) to model an interface VO$_2$ layer with formal 
valency V$^{3.5+}$. 

{\it Ferromagnetism.}-- The large on-site repulsions $U$ and $U'$
suppress the high energy charge fluctuations
$(d^2,d^1)\leftrightarrow(d^3,d^0)$ and only the low energy ones
 $(d^2,d^1)\leftrightarrow(d^1,d^2)$ are allowed. In the paramagnetic
and orbital liquid state all three bands are quarter-filled and
there are no apparent Fermi surface (nesting) related instabilities.
However, there is a potential instability towards FM ordering
because the Hund's coupling favors the high spin state of a $d^2$
configuration and, together with the hopping term, induces the DE
interaction. The critical value of Hund's coupling at which FM
instability appears is estimated from a vanishing determinant of a
two component spin susceptibility for 1D and 2D bands:
\begin{eqnarray}
\left|\!\!\begin{array}{cc}
1 &\!
-4J_{\rm H}{\bar \chi}\\
-2J_{\rm H}\chi&\!1-2J_{\rm H}\chi
\end{array}
\!\!\right|=0~,
\label{stoner}
\end{eqnarray}
where $\chi=\rho(0)/2$ and ${\bar \chi}={\bar \rho}(0)/2$ are uniform
static magnetic susceptibilities for one- and two-dimensional bands,
respectively, $\rho(0)$ and ${\bar \rho}(0)$ are corresponding
density of states at the Fermi level. For an estimate
we set $\rho(0)\simeq 1/W_1$ and ${\bar \rho(0)}\simeq
1/W_2$, where $W_1=4t$ and $W_2=8t$ are bandwidths of 1D and 2D bands,
respectively. This gives a FM instability for
$J_{\rm H}>2(\sqrt{5}-1)t\simeq 2.5 t$.
This inequality is well satisfied for the actual
parameters $J_{\rm H}\simeq 0.7$ eV and $t \simeq 0.2$ eV \cite {Ray07,Note2}.
We thus deal with interacting spinless fermions with
orbital flavors only. In this case the Hamiltonian (\ref{h}) reduces to
\begin{equation}
\label{h1}
H=-\!\!\sum_{i,j,\alpha}t_{ij,\alpha}
d_{i\alpha}^{\dagger}{d}_{j\alpha}
+ \tilde U\!\!\sum_{i,\alpha<\beta}n_{i\alpha}n_{i\beta}+
V\sum_{\langle ij\rangle}n_{i}n_{j}~,
\end{equation}
where $\tilde U =U'-J_{\rm H}=U-3J_{\rm H}$ (using well-known relation
$U'=U-2J_{\rm H}$) is an effective Hubbard repulsion in the
FM state. Optical data in cubic vanadates \cite{Miy02} suggests
the high-spin transition at $\tilde U \sim 2$ eV $\sim 10t$ but we will
consider $\tilde U$ as a free parameter and denote it below simply as $U$.
Each orbital band is half-filled and the corresponding Fermi surfaces 
are fully nested, see Fig.~\ref{fig1}(b). We discuss now the orbital and
charge density waves (ODW and CDW) triggered by such a nesting.

\begin{figure}
\epsfysize=43mm \centerline{\epsffile{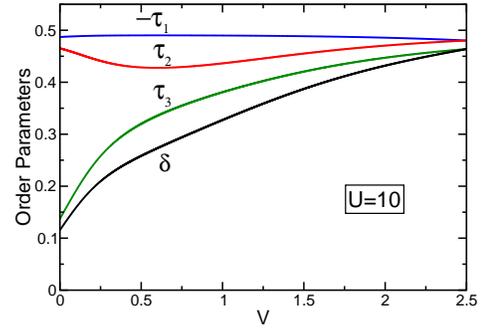}} 
\caption
{(Color online) Order parameters of the orbital ($\tau_{1,2,3}$ for
$d_{yz,xz,xy}$ bands, respectively) and charge ($\delta$) density waves 
vs nearest-neighbor repulsion $V$ at large $U$.}
\label{fig2}
\end{figure}

{\it Orbital and charge density waves.}-- Let us consider the
instabilities towards the ODW and CDW with a modulation wave vector
${\bf Q}=(\pi\, ,\pi)$. We introduce the orbital order parameters
$\tau_{\alpha}$ as $\langle n_{i\alpha}\rangle=n/3+e^{i{\bf Q}{\bf
R}_{i}}\tau_{\alpha}$, where $n=3/2$ is an average electron density. The
corresponding CDW modulation is given by 
$\sum_{\alpha}\langle n_{i\alpha}\rangle - n = e^{i{\bf Q}{\bf
R}_{i}}\delta$, where $\delta=\sum_{\alpha}\tau_{\alpha}$. The order
parameters are calculated within a mean-field approach. We
consider a wide range of $U$ and $V$, in order to see how the
ODW/CDW orderings evolve from a weak coupling regime to the limit of
strong interactions $U\gg t$. While care must be taken in
calculating excitation spectra, the mean-field method gives a reliable
picture of the nature of the ordered phases and of the zero
temperature properties of interest here \cite{Note3a}, and hence is
widely used to study similar problems (including multiorbital
physics at interfaces \cite{Oka04}) even in a regime of strong
correlations \cite{Note3b}. The ground state energy per site is expressed in
terms of order parameters: 
\begin{equation}
E=-\frac{1}{2}\sum_{{\bf
k}\alpha}E_{{\bf k}\alpha}+ \frac{1}{2}U\sum_{\alpha}\tau_{\alpha}^2
+\frac{1}{2}(U-zV)\delta^{2}\, , 
\label{e}
\end{equation}
where $E_{{\bf k}\alpha}= \sqrt{\epsilon_{{\bf k}\alpha}^2+
\Delta_{\alpha}^{2}}$ with
$\Delta_{\alpha}=U\tau_{\alpha}-(U-zV)\delta$, and $z=4$ is a number
of NNs. (The constant contribution
$E_{0}=\frac{1}{3}Un^2+\frac{1}{2}zVn^2$ has been dropped).
Physically, $|\Delta_{\alpha}|$ represent band gaps. The
minimization of the energy $E$ gives the coupled integral equations
$\tau_{\alpha}=\sum_{\bf k}\Delta_{\alpha}/E_{{\bf k}\alpha}$ solved
numerically. For large $U$ and $V$ one finds
$\tau_{3}\alt\tau_{2}=-\tau_{1}\simeq0.5$. A sketch of a fully
saturated  version of the corresponding orbital and charge patterns
is shown in Fig.~\ref{fig1}(c). It consists of a staggered order of
$d_{yz}$ /$d_{xz}$ orbitals, with $d_{xy}$ orbital being
predominantly occupied at one of the sublattices. This results in a
checkerboard charge ordered pattern of $d^2$/$d^1$ states.
\begin{figure}
\epsfysize=43mm \centerline{\epsffile{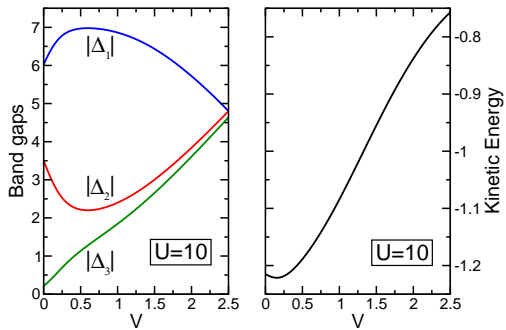}} 
\caption{ (Color
online) Band gaps (left) and kinetic energy per site $K$ (right) vs 
nearest-neighbor repulsion $V$. } 
\label{fig3}
\end{figure}

Shown in Fig.~\ref{fig2} are the ODW and CDW order parameters as a function 
of $V$ at $U=10$ (hereafter, the energy unit $t$ is used). The staggered 
order of $d_{yz}$ and $d_{xz}$ orbitals is strong and weakly affected by 
$V$ (see the $\tau_1$ and $\tau_2$ curves).
The strengths of $d_{xy}$ orbital ($\tau_3$) and charge ($\delta$)
density waves decrease with $V$. Surprisingly, $\delta$ remains finite 
down to $V=0$, {\it i.e.}, a charge modulation is present even in the
case of local interactions only. A similar effect but with different mechanism
has been found within the two-orbital model for manganites \cite{Bri99}. 
Physics behind this unusual picture here is as follows. In the limit 
$U\gg t$ the orbital order parameters $\tau_{1,2}$ for 1D bands are nearly 
saturated and can be expanded in powers of $t/U$. The energy (\ref{e}) is 
then expressed in terms of $\tau_3$ only:
$E=-\frac{1}{4}U-\frac{1}{2}J+2J\tau_{3}^{2}-\frac{1}{2} \sum_{{\bf
k}}\sqrt{\epsilon_{{\bf k}3}^2+(4J\tau_{3})^2}$, where $J=4t^2/U$. The 
minimization of $E$ gives a finite $\tau_{3}$ and non-zero CDW order parameter 
$\delta\simeq\tau_3 - O(J/U)$, because of a singular response of the nested 
$d_{xy}$ Fermi surface at half-filling. 

The orbital and charge density waves with ${\bf Q}=(\pi\, ,\pi)$ modulation
wave vector induce the gaps on the entire Fermi surfaces of all three bands 
and drive the system into the insulating state. In Fig.~\ref{fig3} the band 
gaps are plotted as function of $V$ at $U=10$. At $V=0$, the gaps in the 
$d_{yz}$ and $d_{xz}$ bands ($|\Delta_{1}|$ and $|\Delta_{2}|$, respectively)
are large since they scale as $U$ in the limit $U\gg t$. While the
gap of $d_{xy}$ band is controlled by an effective coupling
constant $\propto J$. In the ``weak-coupling'' limit $J\ll t$
({\it i.e.} strong-coupling $U\gg t$ limit in conventional language)
it is exponentially small, $|\Delta_{3}|\sim t\exp[-t/4J]$ at $V=0$.
At large $V$, the expected $|\Delta_{3}|\sim 2V$ scaling is observed.

\begin{figure}
\epsfysize=43mm \centerline{\epsffile{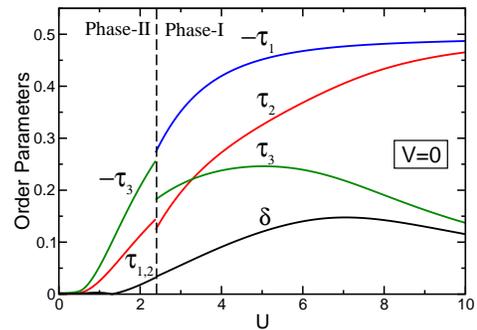}} 
\caption
{(Color online) Order parameters of the orbital ($\tau_{1,2,3}$) and charge
($\delta$) density waves vs $U$. The dashed line marks a first-order
phase transition (see text for details).}
\label{fig4}
\end{figure}

To complete our  analyses, in Fig.~4 the dependence of the order parameters
on $U$ is presented for $V=0$. The dashed line marks a first-order
phase transition at around $U_c\simeq2.4$ from the phase-I,
sketched in Fig.~\ref{fig1}(c), to the phase-II.
In the latter $d_{yz}$ and $d_{xz}$ orbitals predominantly occupy one
sublattice while the density of $d_{xy}$ orbital is higher at the other
one. The ODW order parameters for $d_{yz}$ and $d_{xz}$ bands ($\tau_1$ and
$\tau_2$) are controlled by $U$ and monotonically decrease
with $U$. The non-monotonic behavior of $d_{xy}$ orbital and
CDW order parameters is explained as follows. The effective coupling
constant controlling them vanishes in the limits of small as well
as of large $U$: being of the order of $U$ for $U\ll t$ and
$\propto t^2/U$ for $U\gg t$. For realistic values of model
parameters the phase-I, sketched in Fig.~\ref{fig1}(c), is the
ground state of the system. We emphasize that the system is insulating
for any finite values of $U$ and $V$.

Thus, we arrived at rather unusual situation where DE driven FM and
insulating states coexist \cite{Note4} and, moreover, are closely
interrelated. In fact, the ODW and CDW states are stable only if FM
correlation length is large. In the DE system, a kinetic energy of
electrons defines a stiffness of FM order. Along the same line we
estimate a FM coupling of neighboring $s=1/2$ and $S=1$ spins in
the charge ordered state: $J_{\rm FM}\simeq K/[2S(2s+1)]$, where $K$
is a kinetic energy per site. Fig.~\ref{fig3} shows $K$ as a
function of $V$. Considering a moderate value of $V=2t\simeq 0.4$ eV, we
find $J_{\rm FM}\simeq 40$ meV. This suggests an onset of FM 
correlations at fairly high temperatures and explains insulating 
behavior of VO$_2$ interface in the experiment \cite{Hot07}. 
Further, we predict a transition to a metallic state and large 
magnetoresistivity effects at higher temperature when
FM correlations are reduced. Apart from transport measurements,
magnetic x-ray and optical studies may provide a crucial test for
the theory.

\begin{figure}
\epsfysize=52mm
\centerline{\epsffile{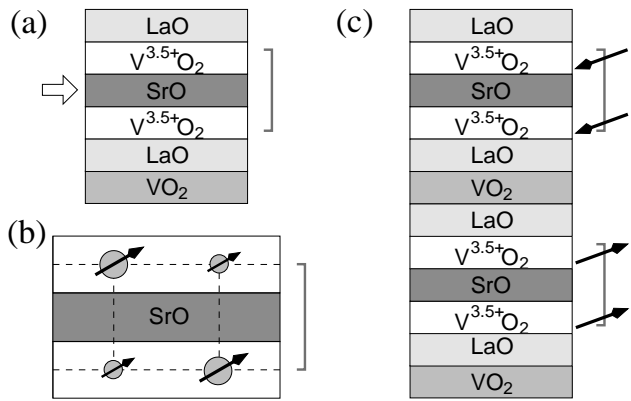}}
\caption{
(a) V$^{3.5+}$O$_2$ bilayer formed by replacement of a LaO (001)
layer of LaVO$_3$ by SrO.
(b) Spin structure of the V$^{3.5+}$O$_2$ bilayer. Large (small) circles
and spins indicate cites where V$^{3+}d_{xy\uparrow}d_{yz\uparrow}$
(V$^{4+}d_{zx\uparrow}$) configuration is favored. The layers are coupled
ferromagnetically via the DE between S=1 and S=1/2 states.
(c) Periodic sequence of V$^{3.5+}$O$_2$ bilayers. Spin coupling through
the intermediate V$^{3+} S=1$ ions of VO$_2$ layer is antiferromagnetic
(as shown here) if the charge ordering patterns of different bilayers
are in-phase.
}
\label{fig5}
\end{figure}

The present work motivates an interesting idea of a superlattice
depicted in Fig.~\ref{fig5}. Here, insertion of SrO planes into LaVO$_3$
can be viewed as a "spatially correlated doping" that generates
{\it two} V$^{3.5+}$O$_2$ planes -- each midway between Sr$^{2+}$O and
La$^{3+}$O layers -- forming a ferromagnetic bilayer. Spins of different
bilayers weakly couple antiferromagnetically (ferromagnetically)
if their CDW-ordering patterns are in-phase (out-of-phase) \cite{Note5}.
Such a direct link between charge and spin structures suggests a magnetic
control of charge sector and vice versa, {\it e.g.}, a relatively weak
magnetic field may lead to the CDW phase-shift. In fact, the proposed
superlattice is similar to the bilayer ruthenates \cite{Cao03} and
manganites \cite{Per98}, which have the same magnetic structure as
in Fig.~\ref{fig5}(c) and show large magnetoresistivity and spin-valve
effects. We notice also that the bilayer coupling may oscillate in sign
as the number of intermediate VO$_2$ planes is varied, provided their
V$^{3+}$ spins stagger along the $c$ axis as in YVO$_3$
at low-temperature \cite{Miy02}.

To conclude, we have studied a Hubbard model for a quarter-filled
$t_{2g}$ bands on a square lattice. Due to a confined geometry at the
interface, at large enough Hund's coupling ($J_{\rm H}\!\agt\!2.5 t$)
the correlated electrons develop a peculiar insulating ferromagnetic
ground state accompanied by the orbital and charge density waves.
This provides a natural explanation for an
insulating behavior of the $p$-type LaVO$_3$/SrTiO$_3$ interface.
The experimental and theoretical studies of superlattices
like in Fig.~\ref{fig5} where a complex electronic reconstruction
takes place coherently over many interfaces remains a future challenge.

We would like to thank B. Keimer, H.Y. Hwang, and P. Horsch for
stimulating discussions. G.J. acknowledges support by GNSF under
the Grant No.06-81-4-100.


\end{document}